\documentstyle[preprint,epsfig,aps]{revtex}

\begin{document}

\title{A new contribution to the nuclear modification factor of non-photonic 
electrons in Au+Au collisions at $\sqrt{s_{NN}} = 200~{\rm GeV}$}

\date{\today}

\author{G.~Mart\'{\i}nez-Garc\'{\i}a$^1$\thanks{Gines.Martinez@subatech.in2p3.fr}, S.~Gadrat$^1$\thanks{Sebastien.Gadrat@subatech.in2p3.fr} and 
P.~Crochet$^2$\thanks{Philippe.Crochet@clermont.in2p3.fr}}

\address{
$^{1}$SUBATECH
(IN2P3/CNRS - Ecole des Mines - Universit\'e de Nantes) Nantes, France\\
$^{2}$Laboratoire de Physique Corpusculaire 
(IN2P3/CNRS - Universit\'{e} Blaise Pascal) Clermont-Ferrand, France}
\maketitle

\begin{abstract}
We investigate the effect of the so-called anomalous baryon/meson enhancement
to the nuclear modification factor of non-photonic electrons in Au+Au 
collisions at $\sqrt{s_{NN}} = 200~{\rm GeV}$.
It is demonstrated that an enhancement of the charm baryon/meson ratio,
as it is observed for non-strange and strange hadrons, 
can be responsible for part of the amplitude of the nuclear modification 
factor of non-photonic electrons.
About half of the measured suppression of non-photonic electrons in the 
$2-4~{\rm GeV/c}$ $p_{\rm t}$ range can be explained by a charm baryon/meson 
enhancement of 5.
This contribution to the non-photonic electron nuclear modification factor 
has nothing to do with heavy quark energy loss.
\end{abstract}

\vspace*{1.cm}


\newpage
One of the most robust experimental evidence for the creation of a new 
state of matter in heavy-ion collisions at the Relativistic Heavy Ion 
Collider (RHIC) is the large suppression of light hadrons at 
high transverse momentum ($p_{\rm t}$)~\cite{RAA_compil}.
This phenomenon is well reproduced by models 
which take into account the radiative energy loss of high $p_{\rm t}$ light 
quarks and gluons propagating through a dense medium of colored quarks and 
gluons~\cite{Salgado:2005pr}.
Further insights into the underlying mechanism can be obtained from the study
of heavy hadrons.
Indeed, heavy hadrons originate from the fragmentation of (heavy) quarks in 
contrast to light hadrons which are predominantly produced by gluons.
Quarks are supposed to lose less energy than gluons in the medium due to a 
smaller color charge coupling.
In addition, radiative energy loss was predicted to be substantially smaller 
for heavy quarks as compared to light quarks because of the so called 
``dead-cone" effect which limits the medium induced radiative energy loss
at forward angles~\cite{Dokshitzer:2001zm}.
Surprisingly, recent data from the PHENIX and the STAR collaborations
in Au+Au collisions at $\sqrt{s_{NN}} = 200~{\rm GeV}$ show that the quenching
of non-photonic electrons lies 
beyond theoretical expectations~\cite{Adare:2006nq,Adler:2005xv,Abelev:2006db}
and is as large as that of light mesons.
This is difficult to understand under the common assumption that non-photonic 
electrons originate predominantly from the decay of $D$ and $B$ mesons, the 
latter originating themselves from heavy ($c$ and $b$) quarks hadronization.
Reconciling these data with model predictions is a real challenge which 
triggers a lot of theoretical activities nowadays.
Only models which assume a very large medium opacity~\cite{Armesto:2005mz} 
or an additional collisional energy 
loss~\cite{Djordjevic:2006kw,vanHees:2005wb}
or a very early fragmentation of heavy quarks in the medium~\cite{Adil:2006ra} 
can describe successfully the data (for a recent review, see~\cite{Ralf}).
It is worth noticing that, in contrast to light hadrons, the heavy flavor 
quenching is, so far, not measured experimentally through identified 
particles, but in an inclusive way via the nuclear modification factor 
($R_{AA}$) of non-photonic electrons\footnote{The nuclear modification factor 
of non-photonic electrons is obtained from the $p_{\rm t}$ distributions of 
non-photonic electrons in $AA$ collisions 
(${\rm d}N^e_{AA}/{\rm d}p_{\rm t}$) and in $pp$ collisions
(${\rm d}N^e_{pp}/{\rm d}p_{\rm t}$) as:
$$R_{\rm AA} = \frac{{\rm d}N^e_{AA}/{\rm d}p_{\rm t}}{<N^{AA}_{\rm coll}>{\rm d}N^e_{pp}/{\rm d}p_{\rm t}}$$
where $<N^{AA}_{\rm coll}>$ is the average number of nucleon-nucleon
collisions corresponding to a given centrality class.}.

In this paper, we point out the possibility that part of the explanation for
the strong suppression of non-photonic electrons might be due to a another
source of electrons namely charmed baryons.
Indeed, whereas light mesons are largely suppressed in heavy ion 
collisions at RHIC, the suppression of non-strange and strange baryons
is observed to be much less in the intermediate $p_{\rm t}$ range
($2 < p_{\rm t} < 4~{\rm GeV/c}$)~\cite{RBM_compil}.
This is commonly referred to as the anomalous baryon/meson enhancement.
It could challenge our understanding of jet quenching at a first glance but
brings however another piece of evidence for the formation of a deconfined
medium.
This anomalous baryon/meson enhancement is well understood in the framework of 
the recombination model which assumes that, at moderate $p_{\rm t}$,
hadronization occurs via the coalescence of ``free" quarks 
(and anti-quarks)~\cite{Fries:2003vb,Fries:2003kq,Greco:2003xt,Greco:2003mm,Hwa:2003ic,Hwa:2002tu}.
In the following, we show that an anomalous baryon/meson enhancement for charm
hadrons leads naturally to a non-photonic electron $R_{AA}$ smaller than
one.
This is essentially due to a smaller semi-leptonic decay branching ratio and 
to a softer decay lepton spectrum of charm baryons as compared to charm mesons.
As a consequence, a fraction of the deviation from the experimentally measured
$R_{AA}$ of non-photonic electrons should not be attributed to energy loss.
After the description of the physical process invoked here, we present simple 
PYTHIA-based simulations to illustrate the magnitude of the effect in 
Au+Au collisions at $\sqrt{s_{NN}} = 200~{\rm GeV}$ and we compare our results 
to the PHENIX data.

The main assumption we put forward is that, in a deconfined medium, charm
baryon production is enhanced relative to charm meson production, as 
compared to the vacuum.
This assumption is qualitatively justified in the framework of the 
recombination model.
Although this model does not provide predictions on charm hadron production 
yet, it successfully describes the baryon/meson enhancement measured in 
Au+Au collisions at $\sqrt{s_{NN}} = 200~{\rm GeV}$.
A relatively good agreement is obtained not only for the light hadron ratio 
$p/\pi^+$,
but also for heavier hadron ratios such as $\Lambda/K^0_s$ and 
$\Omega/\phi$~\cite{Sarah}.
However, extrapolating these results to charm hadrons is not 
straightforward because the mass of the charm quark is much larger than 
that of light and strange quarks.
The consequences are threefold as far as the recombination mechanism is 
concerned.
First, whereas the $p_{\rm t}$ of a light baryon (meson) amounts to 
3 (2) times the initial $p_{\rm t}$ of its valence quarks, the $p_{\rm t}$ of 
a (single) charm baryon or a charm meson is likely to be very close to that of 
the charm quark.
Secondly, considering a light quark and a heavy quark with the same velocity 
(which is the essential requirement for the coalescence 
process to take place~\cite{Lin:2003jy}), 
the heavy quark momentum is much larger than that of light partons.
As a consequence, one can expect the enhancement of charm baryon/meson 
to appear at higher $p_{\rm t}$. 
The recombination model indeed predicts, for non-strange and strange
hadrons, that the heavier the hadron, the larger the $p_{\rm t}$ of the 
baryon/meson enhancement~\cite{Sarah}.
Finally, the fragmentation time of heavy quarks is small as compared
to light quarks. According to~\cite{Adil:2006ra}, the formation time of 
a $10~{\rm GeV/c}$ pion, $D$ meson and $B$ meson is $20$, $1.5$ and 
$0.4~{\rm fm/c}$, respectively and it is as small as 
$\sim 3~{\rm fm/c}$ for a $\Lambda_c$ with $p_{\rm t} = 20-30~{\rm GeV/c}$.
Due to these considerations, it is obvious that the baryon/meson enhancement
for non-charm hadrons and charm hadrons can be significantly different.
We stress that our aim is neither to investigate these differences nor 
to quantify the magnitude of the charm baryon/meson enhancement.
This would require a sophisticated theoretical investigation which goes 
clearly beyond the scope of this paper and, to our knowledge,
no information on this topic is available in the literature yet.
Therefore, in the following, we only assume that, in view of experimental 
results on the baryon/meson enhancement for non-strange and strange hadrons, 
a similar enhancement is a priori conceivable for charm hadrons. 
Remarkably, such an enhancement has strong implications on the 
nuclear modification factor of non-photonic electrons.
It leads to a decrease of the yield of non-photonic electrons in 
$AA$ collisions because, as shown in Tab.~\ref{tab_PDG}, the inclusive 
semi-leptonic decay branching ratio of charm baryons is smaller than that of 
charm mesons (and the hadronization process does not change the total charm 
quark number).
Therefore, the nuclear modification factor of non-photonic electrons should 
decrease as well.
This can be easily illustrated in the following way.
Assuming that a $AA$ collision is a perfect sum of
$pp$ collisions (i.e. $R_{AA} = 1$) and that the relative yields 
of $D$ mesons are the same in $pp$ and in $AA$ collisions, 
a $p_{\rm t}$ integrated $R_{AA}$ 
can be calculated for different $\Lambda_c/D$ enhancement 
factors ${\cal C} = N_{\Lambda_c,\bar{\Lambda_c}} / N_D = 
(N_{\Lambda_c} + N_{\bar{\Lambda_c}}) / (N_{D^+} + N_{D^-} + 
N_{D^0} + N_{\bar{D^0}} + N_{D^+_s} + N_{D^-_s})$
according to 
\begin{equation}
R_{AA} = 	
\frac{1 + N_{\Lambda_c,\bar{\Lambda_c}}/N_D }
{1 + {\cal C} (N_{\Lambda_c,\bar{\Lambda_c}}/N_D)}
	\frac{1 + {\cal C} (N_{\Lambda_c\rightarrow e} / N_{D\rightarrow e})}
	{1 + N_{\Lambda_c\rightarrow e} / N_{D\rightarrow e}}
\label{eq1}
\end{equation}
where 
\begin{equation}
N_{\Lambda_c\rightarrow e} / N_{D\rightarrow e} =  
      \frac{(N_{\Lambda_c,\bar{\Lambda_c}}/N_D)BR_{\Lambda_c,\bar{\Lambda_c}}}
           {(N_{D^\pm}/N_D)BR_{D^\pm}+(N_{D^0,\bar{D^0}}/N_D)BR_{D^0,\bar{D^0}}+(N_{D_s^\pm}/N_D)BR_{D^\pm_s}}.
\label{eq2}
\end{equation}
$N$ is the charm hadron yield and $BR$ is the hadron 
semi-leptonic decay branching ratio.
According to Tab.~\ref{tab_PDG}, $N_{\Lambda_c,\bar{\Lambda_c}}/N_D=7.3\%$, 
$N_{D^\pm}/N_D=21\%$, $N_{D^0,\bar{D^0}}/N_D=67\%$, and $N_{D_s^\pm}/N_D=12\%$
such that $N_{\Lambda_c\rightarrow e} / N_{D\rightarrow e} = 3.63\%$.
Enhancement factors ${\cal C}$ of 5 and 12 therefore lead to a $R_{AA}$ 
of 0.90 and 0.79 respectively.
It is shown below that the effect is further amplified in the intermediate
$p_{\rm t}$ range as a result of a softer decay electron spectrum of
$\Lambda_c$ as compared to that of $D$ mesons.

\begin{figure}[htb]
 \begin{center}
  \epsfig{file=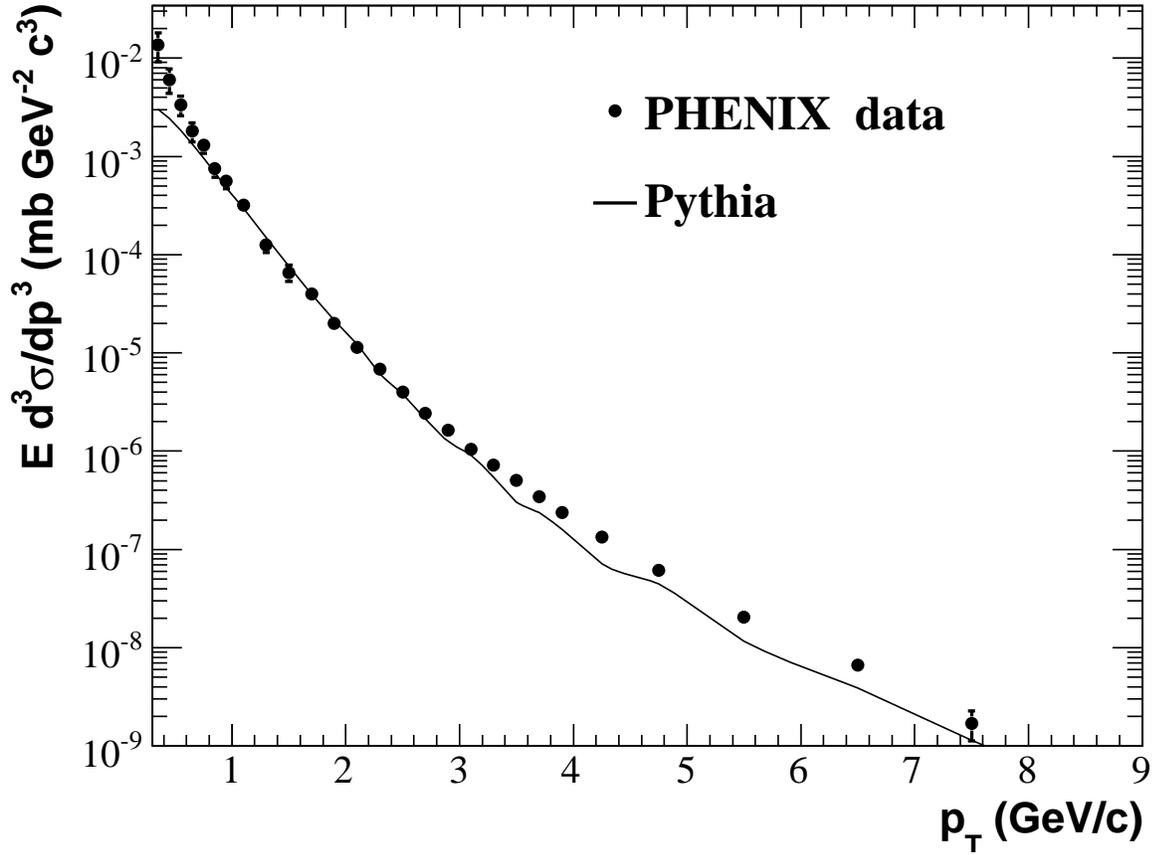,width=1.\linewidth}
 \end{center}
 \caption{Invariant differential cross-section of non-photonic electrons (dots)
	measured in $pp$ collisions at 
	$\sqrt{s} = 200~{\rm GeV}$~\protect\cite{NewPhenix}.
	The solid curve shows the result of the PYTHIA simulation as described
	in the text. 
	The simulated spectrum is normalized from the integration of the 
	measured spectrum in the 
	range ${\rm 1.4} < p_{\rm t} < {\rm 4}~{\rm GeV/c}$.}
 \label{fig_eldndpt} 
\end{figure}

\begin{figure}[htb]
 \begin{center}
  \epsfig{file=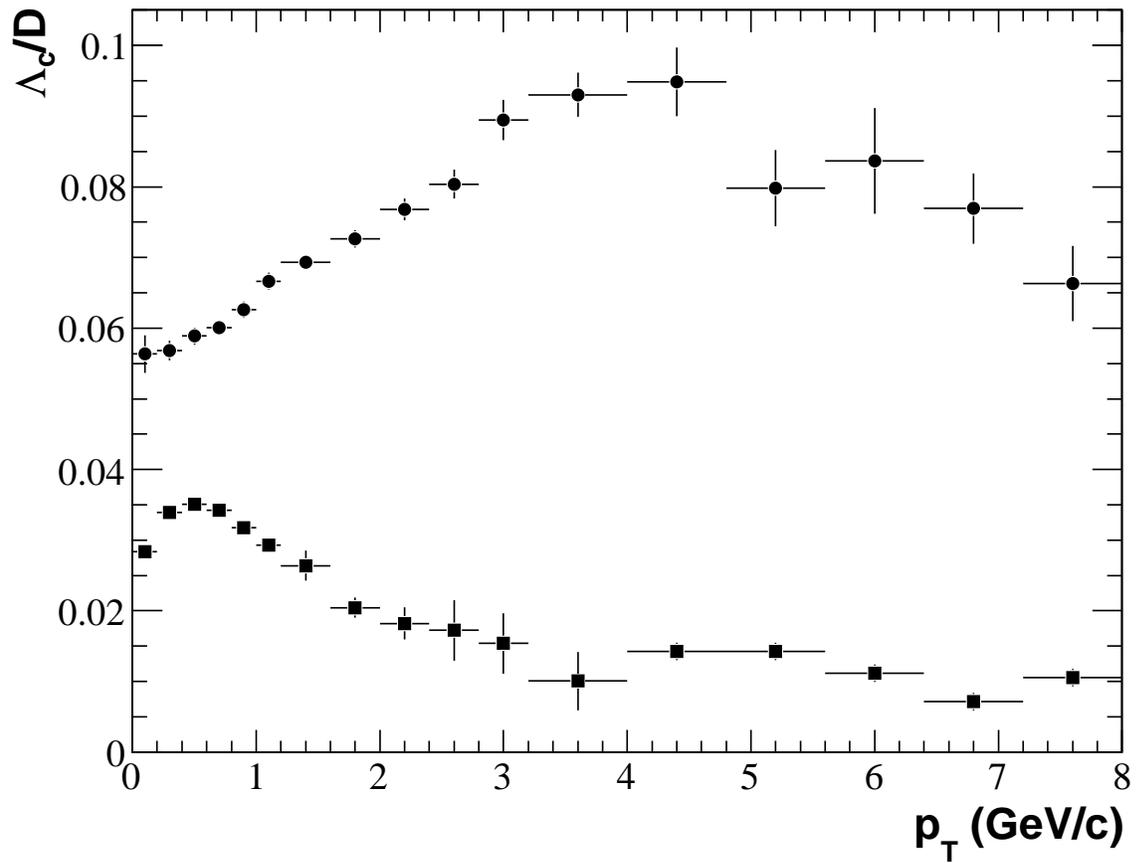,width=1.\linewidth}
 \end{center}
 \caption{Charm baryons over charm mesons (dots) and 
	decay electrons from charm baryons over decay 
	electrons from charm mesons (squares) versus $p_{\rm t}$, from 
	the PYTHIA 
	simulation described in the text.}
\label{fig_dratios} 
\end{figure}

In order to investigate this effect in more details, we have performed simple 
simulations with the PYTHIA-6.152~\cite{Sjostrand:1993yb} 
event generator.
The PYTHIA input parameters were first tuned according to~\cite{Adcox:2002cg} 
and the PHENIX acceptance cut ($|\eta| < 0.35$) was applied in order to 
correctly reproduce
the $p_{\rm t}$ distribution of non-photonic electrons measured in $pp$
collisions at $\sqrt{s} = 200~{\rm GeV}$~\cite{NewPhenix}.
As it can be seen from Fig.~\ref{fig_eldndpt}, the agreement between the 
simulation and the data is rather good except at very low $p_{\rm t}$ and 
in the high $p_{\rm t}$ region where the simulation under-predicts the data.
The disagreement in the high $p_{\rm t}$ region is due to the fact that we 
consider, in the simulation,
only electrons from charm decay and neglect electrons from bottom decay 
which are expected to
dominate the spectrum for $p_{\rm t} \gtrsim 4~{\rm GeV/c}$~\cite{NewPhenix}.
On the other hand, it can be observed from Tab.~\ref{tab_PDG} that the 
$\Lambda_c/D$ ratio 
amounts to 7.3\% (in $4\pi$) and, as mentionned above, to 3.63\% after 
convolution of the species yields with their corresponding semi-leptonic 
decay branching ratio.
Figure~\ref{fig_dratios} shows that, in the $2-4~{\rm GeV/c}$ $p_{\rm t}$ 
region of interest discussed hereafter, this ratio is even smaller 
($\sim 1.5\%$) because the decay electron spectrum of 
$\Lambda_c$ is softer than that of $D$ mesons.
In this respect, an enhancement of the $\Lambda_c/D$ ratio in central heavy 
ion collisions induces two different mechanisms of non-photonic electron 
yield suppression:
i) integrated yields are suppressed due to the lower semi-electronic branching 
ratio of the $\Lambda_c$ 
and ii) the softer electron $p_{\rm t}$ distribution from 
$\Lambda_c$ decay leads to an additional suppression at
intermediate $p_{\rm t}$.

\begin{figure}[htb]
 \begin{center}
  \epsfig{file=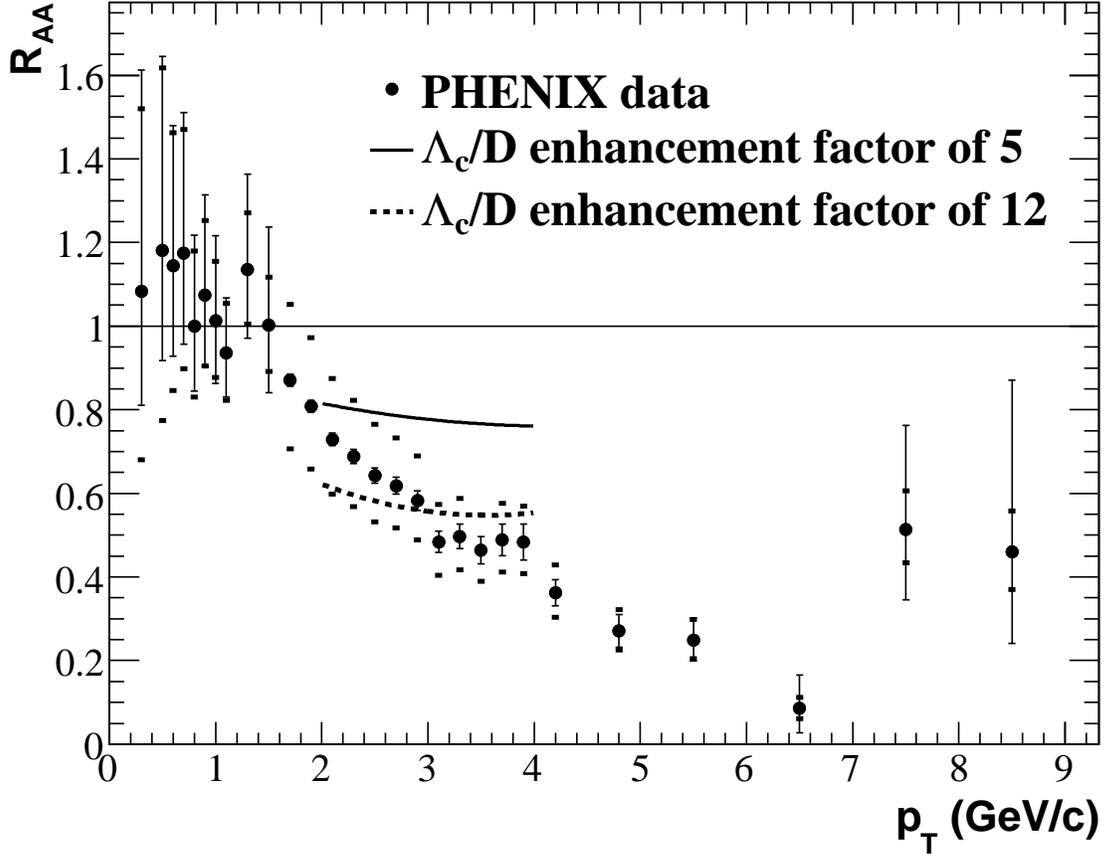,width=1.\linewidth}
 \end{center}
 \caption{Nuclear modification factor of non-photonic electrons (dots) 
	measured in central (0-10\%) Au+Au collisions at 
	$\sqrt{s_{NN}} = 200~{\rm GeV}$~\protect\cite{Adare:2006nq}.
	The solid and dotted curves correspond to the results of the 
	simulation described in the text for a $\Lambda_c/D$ enhancement 
	factor of 5 and 12 respectively.}
 \label{fig_RAA} 
\end{figure}

We then re-evaluate the non-photonic electron decay spectrum after introducing 
a $\Lambda_c/D$ enhancement chosen to 5 and 12.
This enhancement is assumed to be flat in $p_{\rm t}$ and 
it is applied such that the $p_{\rm t}$-differential charm cross-section 
is conserved.
The latter is an arbitrary choice that could be justified since most of the 
charm hadron transverse momentum 
is given by the charm quark whatever, baryon or meson, this hadron is.
We finally compute the $R_{AA}$ ratio from the previous non-photonic electron
$p_{\rm t}$ spectra neglecting all effects like shadowing or quenching
but assuming that the only medium induced effect is the $\Lambda_c/D$ 
enhancement.
The nuclear modification factor is therefore simply obtained from the ratio
of the non-photonic electron $p_{\rm t}$ distribution with 
a $\Lambda_c/D$ enhancement to the original distribution.
By doing so, one implicitly takes into account the $p_{\rm t}$ dependence 
of the $\Lambda_c/D$ ratio which is not flat (Fig.~\ref{fig_dratios}).
The results are shown in Fig.~\ref{fig_RAA} together with the PHENIX data.
The simulated $R_{AA}$ ratio is shown only for
$2 < p_{\rm t} < 4~{\rm GeV/c}$ since shadowing and electrons from
bottom decay, which play an important role at lower and higher $p_{\rm t}$
respectively, are not considered in the simulation.
One can see from Fig.~\ref{fig_RAA} that a $\Lambda_c/D$
enhancement of 5 can already explain $\sim 50\%$ of the suppression of 
non-photonic electrons in the $2-4~{\rm GeV/c}$ $p_{\rm t}$ range.
For illustration, we show that, in the extreme 
scenario of an enhancement of 12, the data can be 
satisfactorily described without invoking any heavy-quark energy loss.
The shape of the simulated $R_{AA}$ is obviously determined by the $p_{\rm t}$ 
distribution of $\Lambda_c/D$ enhancement that we have assumed to be flat.
Note that the values of $R_{AA}$ at $p_{\rm t} \sim 3~{\rm GeV/c}$ can be 
roughly
obtained as well from Fig.~\ref{fig_dratios} and Eq.~\ref{eq1} and~\ref{eq2}, 
assuming that the decay electrons at $p_{\rm t} \sim 3~{\rm GeV/c}$ originate 
from charm hadrons with $p_{\rm t} \gtrsim 4~{\rm GeV/c}$.

In summary, we have shown that an enhancement of the $\Lambda_c/D$ ratio 
in heavy ion collisions has dramatic consequences on the nuclear modification 
factor of non-photonic electrons.
Such an enhancement, which is observed for non-strange and strange 
hadrons, would significantly lower the $R_{AA}$ of non-photonic electrons 
(or muons) at intermediate $p_{\rm t}$.
This results from
i) the semi-leptonic decay branching ratio of charm baryons which 
is smaller than that of charm mesons 
and ii) the decay lepton spectrum from charm baryons which is softer than that 
of charm mesons.
We conclude that it is therefore premature to interpret the non-photonic 
electron $R_{AA}$ data before a possible enhancement of the $\Lambda_c/D$ ratio
is measured experimentally and/or investigated theoretically.
Heavy quark energy loss can be studied in a much cleaner way from the nuclear 
modification factor of exclusively reconstructed charm hadrons.
Such measurements should be possible with the RHIC-II
experiments~\cite{jacak} and with the ALICE experiment at the 
LHC~\cite{Alessandro:2006yt}.
We finally note that the $\Lambda_c/D$ enhancement can possibly influence
the elliptic flow of non-photonic electrons as well.

\section*{Note}
After the publication of the present work in arXiv we have realized
that a similar work has already been published by P.~Sorensen and 
X.~Dong in PRC 74 (2006) 024902.
However, the assumptions made in the two approaches are different 
and lead to different conclusions. 
In PRC 74 (2006) 024902, assuming a $\Lambda_c/D$ ratio of $1.7$, similar to
the measured $\Lambda/K^0_s$ ratio, about 35\% of the 
of non-photonic electron $R_{AA}$ data is reproduced.
In our approach, the $R_{AA}$ data are reproduced 
by using the $\Lambda_c$ and $D$ spectra from PYTHIA and assuming a 
$\Lambda_c/D$ ratio of $\sim 1$ (i.e $12 \times 0.08$).

\acknowledgements
We gratefully acknowledge Anton Andronic, Pol-Bernard Gossiaux 
and St\'ephane Peign\'e for 
carefully reading the manuscript and for valuable suggestions.

\begin{center}
\begin{table}[h]
\caption{Inclusive decay branching ratio ($BR$) of charm hadrons
into $e$ $+$ anything~\protect\cite{Yao:2006px} and yield ($N$)
of charm hadrons (in 4$\pi$) in $pp$ collisions at $\sqrt{s} = 200~{\rm GeV}$ 
from the present PYTHIA simulation using input parameters as described 
in~\protect\cite{Adcox:2002cg}.
The total cross-section for charm production is normalized to the 
experimental value obtained in~\protect\cite{Adcox:2002cg}.
$N_{\Lambda_c}$ and $N_{\bar{\Lambda_c}}$ include primarily produced 
$\Lambda_c$ and $\bar{\Lambda_c}$ as well as those from $\Sigma_c$ and 
$\bar{\Sigma_c}$ decay.}
\begin{tabular}{lllllllll}
Hadron        & $D^+$ & $D^-$ & $D^0$ & $\bar{D^0}$ & $D^+_s$ & $D^-_s$ &
$\Lambda_c$ & $\bar{\Lambda_c}$ \\
\hline
$BR$ (\%) & \multicolumn{2}{c}{$17.2\pm 1.9$} & 
	  \multicolumn{2}{c}{$6.71\pm 0.29$} & 
          \multicolumn{2}{c}{$8^{+6}_{-5}$} & 
          \multicolumn{2}{c}{$4.5\pm 1.7$}\\ 
\hline 
$N$ $(\times 10^{-3}$) & 3.00 & 3.07 & 9.31 & 9.85 & 1.82 & 1.60 & 1.23 & 0.85 \\
\end{tabular}
\label{tab_PDG}
\end{table}
\end{center}

\newpage

\newpage

\newpage


\newpage

\end{document}